\documentclass[10pt,conference]{IEEEtran}
\usepackage{graphicx} % support the \includegraphics command and options
\usepackage{color}
\usepackage{psfrag}
\usepackage{epsfig}

\definecolor{Red}{rgb}{1,0,0}
\definecolor{Blue}{rgb}{0,0,1}
\definecolor{Olive}{rgb}{0.41,0.55,0.13}
\definecolor{Green}{rgb}{0,1,0}
\definecolor{MGreen}{rgb}{0,0.8,0}
\definecolor{DGreen}{rgb}{0,0.55,0}
\definecolor{Yellow}{rgb}{1,1,0}
\definecolor{Cyan}{rgb}{0,1,1}
\definecolor{Magenta}{rgb}{1,0,1}
\definecolor{Orange}{rgb}{1,.5,0}
\definecolor{Violet}{rgb}{.5,0,.5}
\definecolor{Purple}{rgb}{.75,0,.25}
\definecolor{Brown}{rgb}{.75,.5,.25}
\definecolor{Grey}{rgb}{.5,.5,.5}

\usepackage{booktabs} % for much better looking tables
\usepackage{array} % for better arrays (eg matrices) in maths
\usepackage{paralist} % very flexible & customisable lists (eg. enumerate/itemize, etc.)
\usepackage{verbatim} % adds environment for commenting out blocks of text & for better verbatim
\usepackage{subfigure} % make it possible to include more than one captioned figure/table in a single float
\usepackage{amsmath,amssymb}

\newtheorem{corollary}{Corollary}
\newtheorem{claim}{Claim}
\newtheorem{lemma}{Lemma}

\newtheorem{bound}{Bound}

\newtheorem{remark}{Remark}

\newcommand{\p}{{\rm P}}

\def\cA{{\cal A}}
\def\cB{{\cal B}}
\def\cC{{\cal C}}

\def\cU{{\cal U}}

\def\cY{{\cal Y}}

\begin{document}

% paper title
\title{On  3-receiver broadcast channels 
with 2-degraded message sets}

% author names and affiliations
% use a multiple column layout for up to three different
% affiliations
\author{
\authorblockN{Chandra Nair}
\authorblockA{Department of Information Engineering\\
Chinese University of Hong Kong \\
Sha Tin, N.T., Hong Kong\\
Email: chandra@ie.cuhk.edu.hk}
\and
\authorblockN{Zizhou Vincent Wang}
\authorblockA{Department of Information Engineering\\
Chinese University of Hong Kong \\
Sha Tin, N.T., Hong Kong\\
Email: zzwang6@ie.cuhk.edu.hk}
% Email: medard@mit.edu}
}

% make the title area
\maketitle

\maketitle

\begin{abstract}
We consider a broadcast channel with 3 receivers  and 2 messages $(M_0, M_1)$ where two of the three receivers need to decode messages $(M_0, M_1)$ while the remaining one just needs to decode the message $M_0$. We study the best known inner and outer bounds under this setting, in an attempt to find the deficiencies with the current techniques of establishing the bounds.  We produce a simple example where we are able to explicitly evaluate the inner bound and show that it differs from the general outer bound. For a class of channels where the general inner and outer bounds differ, we use a new argument to show that the inner bound is tight.
\end{abstract}

\section{Introduction}

The broadcast channel with degraded message sets was initially studied by K\'{o}rner and M\"{a}rton \cite{kom77} for two receivers and more recently in \cite{dit06,pdt07,nae07b} for three and more receivers. K\'{o}rner and M\"{a}rton \cite{kom77} established the capacity region for the degraded message sets with two receivers and some capacity regions for three or more receivers were established  \cite{dit06,pdt07} by showing that the straightforward extension of the inner bound in \cite{kom77} was optimal. In \cite{nae07b}, an idea called {\em indirect decoding} was introduced and the authors showed that this could be used to enhance (in some cases strictly) the straightforward extension of the inner bound by K\'{o}rner and M\"{a}rton. Unfortunately, the new inner bounds \cite{nae07b} become quite messy and  unwieldy due to the introduction of many auxiliary random variables. However there is still one class of broadcast channels with degraded message sets where the idea of indirect decoding does not yield any region better than the straightforward extension of the K\'{o}rner and M\"{a}rton inner bound, and this is the scenario of interest here. 

Consider a 3-receiver broadcast channel with 2 messages $(M_0,M_1)$ with the following decoding requirement. Receivers $Y_1$ and $Y_2$ need to decode both messages $(M_0,M_1)$ while receiver
$Y_3$ needs to decode only message $M_0$. The traditional inner and outer bounds presented below remain the best known inner and outer bounds for this class of broadcast channels. In this paper we look at the general inner and outer bounds for this scenario in a greater detail. We show that these bounds differ in general, and that there is a class of channels where the inner bound is tight and the outer bound is weak.

There are two main contributions in this paper: the first one is the technique (same spirit as Mrs. Gerber's lemma \cite{wyz73}) used to evaluate the boundary of a particular inner bound; the second is the use of a ($\frac 12 + \epsilon$)-codebook\footnote{An $\eta$-codebook is used to denote a codebook whose probability of error is bounded above by $\eta$.} rather than an $\epsilon$-codebook to establish the capacity region.

\begin{bound}
 \label{bd:inner}
The union of the following set of rate pairs $(R_0,R_1)$ satisfying
\begin{align*}
 R_0 &\leq I(U;Y_3)\\
 R_1 &\leq \min\{I(X;Y_1|U),I(X;Y_2|U)\}\\
 R_0 + R_1 &\leq \min\{I(X;Y_1),I(X;Y_2)\}
\end{align*}
over all pairs of random variables $(U,X)$ such that $U \to X \to (Y_1,Y_2,Y_3)$ forms a Markov chain constitutes an inner bound to the capacity region.
\end{bound}

\begin{bound}
 \label{bd:outer}
The union over the set of rate pairs $(R_0,R_1)$ satisfying
\begin{align*}
 R_0 &\leq \min\{I(U_1;Y_3),I(U_2;Y_3)\}\\
 R_0+R_1 &\leq \min\{I(U_1;Y_3)+I(X;Y_1|U_1), \\
 & \qquad I(U_2;Y_3)+I(X;Y_2|U_2)\}\\
 R_0 + R_1 &\leq \min\{I(X;Y_1),I(X;Y_2)\}
\end{align*}
over all possible choices of random variables $(U_1,U_2,X)$ such that $(U_1,U_2) \to X \to (Y_1,Y_2,Y_3)$ forms a Markov chain
constitutes an outer bound for this channel.
\end{bound}

The above bounds are traditional, i.e. can be obtained using standard techniques. The inner bound is a straightforward extension of the achievability argument in \cite{kom77} and the outer bound can be deduced by arguments in \cite{elg79, nae07}, etc.

\begin{remark}
 It is also possible to include the constraint 
 $$R_0 \leq \min \{I(U_1;Y_1), I(U_2;Y_2) \}$$ into the outer bound. However, it is quite straightforward to show that the region obtained by adding this inequality is identical to the bound we presented.
\end{remark}

These bounds are known to be tight in all of the following special cases,
\begin{itemize}
 \item Receiver $Y_1$ is a {\em less noisy} receiver than $Y_3$ and $Y_2$ is a {\em less noisy} receiver than $Y_3$ \cite{dit06,nae07b},
 \item $Y_3$ is a deterministic function of $X$,
 \item $Y_1$ is a {\em more capable} receiver than $Y_2$ (or vice-versa),
 \item $Y_3$ is a {\em more capable} receiver than $Y_2$ (or $Y_1$),
\end{itemize}

The last two cases are very straightforward and the proof is omitted. When $Y_3$ is a deterministic function of $X$, note that it is not difficult to show that by taking the convex closure of the regions obtained by setting $(i)$ $U=Y_3$ and $(ii)$ $U=\emptyset$ in  the inner bound exhausts the following region,
\begin{align*}
R_0 &\leq H(Y_3) \\
R_0 + R_1 &\leq \min\{I(X;Y_1),I(X;Y_2)\}
\end{align*}
and this clearly forms an outer bound to the capacity region.

One class of channels that does not fall into any of the cases is the following channel shown in Figure \ref{fig:bsscbsc} below.
The channel $X \to (Y_1,Y_2)$ represents a {\em binary skew-symmetric} (BSSC) broadcast channel \cite{nae07,hap79} and the channel $X \to Y_3$ represents 
a {\em binary symmetric} (BSC) with crossover probability $p$, with $0 \leq p \leq \frac 12$.

\begin{figure}[ht]
\begin{center}
\begin{psfrags}
\psfrag{a}[r]{$0$}
\psfrag{b}[r]{$1$}
\psfrag{c}[r]{$0$}
\psfrag{d}[r]{$1$}
\psfrag{e}[r]{$0$}
\psfrag{f}[r]{$1$}
\psfrag{g}[r]{$0$}
\psfrag{h}[r]{$1$}
\psfrag{i}[r]{$X$}
\psfrag{j}[l]{$Y_1$}
\psfrag{k}[l]{$Y_2$}
\psfrag{l}[l]{$Y_3$}
\psfrag{m}[r]{$\frac 12$}
\psfrag{n}[r]{$\frac 12$}
\psfrag{o}[r]{$p$}
\includegraphics[width=0.75\linewidth,angle=0]{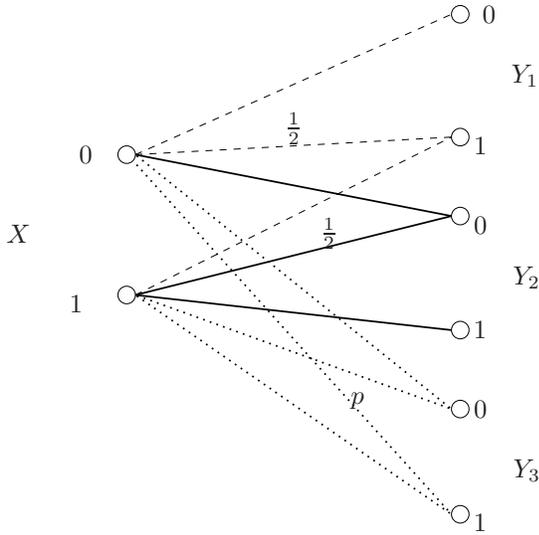}
\end{psfrags}
\caption{3-receiver broadcast channel}
\label{fig:bsscbsc}
\end{center}
\end{figure}

In the next section we evaluate Bound \ref{bd:inner} for this channel.  Based on the symmetry, it is very natural to believe that the auxiliary channel $U \to X$ must be a BSC with some cross over probability $s$. In the next section, we prove that this is indeed the case. This uses a technique similar in spirit to Wyner and Ziv's technique of using Mrs. Gerber's lemma \cite{wyz73}. We will also show that the Bound \ref{bd:outer} yields a strictly larger region for this channel. Finally, we will show that the region represented by Bound \ref{bd:inner} constitutes the {\em capacity region} for this channel.

\section{Evaluation of the inner bound}

In the evaluation of the inner bound, we divide the range $0 \leq p \leq \frac{1}{2}$ into two regions, $0 \leq p \leq p_{max}$ and $p_{max} \leq p \leq \frac{1}{2}$, where $p_{max} \in [0,\frac 12]$ is the unique solution of 
$$ 1-h(p) = h(\frac 14) - \frac 12;$$
i.e. the value of $p$ at which capacity of the BSC  matches the term $\max_{p(x)} \min\{I(X;Y_1),I(X;Y_2)\}$. The numerical value of $p_{max} \approx 0.184$.

\subsection{Evaluation of the inner bound, $ 0 \leq p \leq p_{max}$}
In the region $0 \leq p \leq p_{max}$ it is straightforward to see that the inner bound reduces to the following region (obtained via a time-division between the two auxiliary channels: $(i)$ $U=\emptyset$ and $(ii)$ $U=X$, and in each case, setting $\p(X=0)=0.5$),
$$R_0 + R_1 \leq h(\frac 14) - \frac 12,$$
which clearly matches the outer bound (Bound \ref{bd:outer}). {\em Thus for $0 \leq p \leq p_{max}\approx 0.184 $, the inner and outer bounds are tight and give the capacity region.}
 
\subsection{Evaluation of the inner bound, $p_{max} \leq p \leq \frac{1}{2}$}

Let $\cU=\{1,2,...,m\}$ and let $\p(U=i) = u_i$ and $\p(X=0|U=i)=s_i$. Further, let 
$$h(x) = -x \log_2x - (1-x) \log_2(1-x) $$ 
denote the binary entropy function. 

Using these notations we have,
\begin{align*}
 I(U;Y_3) &= h(\sum_i u_i (s_i (1-p) + (1-s_i)p) ) \\
 &\qquad - \sum_i u_i h(s_i (1-p) + (1-s_i)p) \\
 I(X;Y_1|U) &= \sum_i u_i h(\frac{s_i}{2}) - \sum_i u_i s_i \\
 I(X;Y_2|U) &= \sum_i u_i h(\frac{1-s_i}{2}) - \sum_i u_i (1-s_i) \\
 I(X;Y_1) &= h(\sum_i  \frac{u_i s_i}{2} ) - \sum_i u_i s_i \\
 I(X;Y_2) &= h(\sum_i  \frac{u_i (1-s_i)}{2} ) - \sum_i u_i (1-s_i).
\end{align*}

Define $\tilde{\cU} = \{1,2,...,m\} \times \{1,2\}$, $\p(\tilde{U}=(i,1)) = \frac{u_i}{2}$,
$\p(X=0|\tilde{U}=(i,1))=s_i$, $\p(\tilde{U}=(i,2)) = \frac{u_i}{2}$, and
$\p(X=0|\tilde{U}=(i,2))=1-s_i$. This induces an $\tilde{X}$ with $P(\tilde{X}=0)=\frac{1}{2}$. It is straightforward
to see the following:
\begin{align*}
 I(\tilde{U};\tilde{Y}_3) &\geq I(U;Y_3) \\
 I(\tilde{X};\tilde{Y}_1|\tilde{U}) & = I(\tilde{X};\tilde{Y}_2|\tilde{U}) = \frac{1}{2}(I(X;Y_1|U) + I(X;Y_2|U)) \\
 & \geq \min\{I(X;Y_1|U), I(X;Y_2|U)\} \\
 I(\tilde{X};\tilde{Y}_1) &= I(\tilde{X};\tilde{Y}_2) \geq \frac{1}{2}(I(X;Y_1) + I(X;Y_2))
\end{align*}

From this it follows that for every $U$ replacing $U$ by $\tilde{U}$ leads to a larger achievable region. Hence to
evaluate Bound \ref{bd:inner}, it suffices to maximize over all auxiliary random variables 
of the form $U$ defined by: $\cU = \{1,2,...,m\} \times \{1,2\}$, $\p(U=(i,1)) = \frac{u_i}{2}$,
$\p(X=0|U=(i,1))=s_i$, $\p(U=(i,2)) = \frac{u_i}{2}$, and
$\p(X=0|U=(i,2))=1-s_i$.

Under this notation we have the following expression for the rate region given in Bound \ref{bd:inner},
\begin{align*}
 R_0 &\leq I(U;Y_3) \\
 &= h\big(\frac 12\big) - \sum_i u_i h(s_i (1-p) + (1-s_i)p), \\
 R_1 & \leq \min\{I(X;Y_1|U), I(X;Y_2|U)\} \\
 &=  \sum_i \frac{u_i}{2} \left( h(\frac{s_i}{2}) + h(\frac{1-s_i}{2})\right) - \frac 12, \\
R_0 + R_1 &\leq \min\{I(X;Y_1),I(X;Y_2)\} \\
&= h\big(\frac 14\big) - \frac 12.
\end{align*}
Using the symmetry of the function $h(x)=h(1-x)$ we note that
$$h(s_i (1-p) + (1-s_i)p) = h((1-s_i) (1-p) + s_ip)$$
and thus the above region is constant under the transformation $s_i \to 1-s_i$, implying we can 
restrict $s_i$ to take values only in $0 \leq s_i \leq \frac 12$.

Before we proceed to determine the boundary of this region, we prove the following lemma. 

\subsection{An inequality for a class of functions}

\begin{lemma}
\label{le:newineq}
Let $f(x)$ and $g(x)$ be two non-negative  and strictly increasing functions that are differentiable in the region $x \in [x_1,x_2]$. Further assume that $\frac{f^{(1)}(x)}{g^{(1)}(x)}$ is a decreasing function, where $f^{(1)}(x)$ and $g^{(1)}(x)$ denote the derivatives of the function. Given any $u, 0 \leq u \leq 1$, let $x_{int}$ be uniquely defined according to $ f(x_{int}) = u f(x_1) + (1-u) f(x_2). $
Then the following holds,
\begin{equation*}
  g(x_{int}) \leq u g(x_1) + (1-u)g(x_2). 
\end{equation*}
\end{lemma}

\medskip

\begin{proof}
We have $u(f(x_{int}) - f(x_1)) = (1-u) (f(x_2) - f(x_{int}))$, and we wish to show that $u(g(x_{int}) - g(x_1)) \leq (1-u) (g(x_2) - g(x_{int}))$. Since all the terms are positive, this reduces to showing
$$ \frac{f(x_{int}) - f(x_1)}{g(x_{int}) - g(x_1)} \geq \frac{f(x_2) - f(x_{int})}{g(x_2) - g(x_{int})}. $$ However, this is immediate as shown below.

From the fact that $\frac{f^{(1)}(x)}{g^{(1)}(x)}$ is a decreasing function, we have
$$ \frac{\int_{x_1}^{x_{int}} f^{(1)}(x) dx }{\int_{x_1}^{x_{int}} g^{(1)}(x) dx }
\geq \frac{f^{(1)}(x_{int})}{g^{(1)}(x_{int})} \geq \frac{\int_{x_{int}}^{x_{2}} f^{(1)}(x) dx }{\int_{x_{int}}^{x_{2}} g^{(1)}(x) dx }$$
\end{proof}

Repeated applications of Lemma \ref{le:newineq} leads to the following corollary - potentially of independent interest.

\begin{corollary}
\label{co:ineq}
Let $f(x)$ and $g(x)$ be two non-negative and strictly increasing functions that are differentiable in the region $x \in [x_1,x_2]$. Further assume that $\frac{f^{(1)}(x)}{g^{(1)}(x)}$ is a decreasing function, where as before $f^{(1)}(x)$ and $g^{(1)}(x)$ denote the derivatives of the function. Given any $u_i \geq 0, \sum_i u_i = 1$, and $y_i \in [x_1,x_2]$, let $x_{int}$ be uniquely defined according to $ f(x_{int}) = 
\sum_i u_i f(y_i). $ Then the following holds
$$g(x_{int}) \leq \sum_i u_i g(y_i).$$
\end{corollary}

\subsection{Determining the boundary rate pairs}

We use the Corollary \ref{co:ineq} to determine the boundary of the region. We make the following identifications,
let $f(x) = h(\frac{x}{2}) + h(\frac{1-x}{2}) - 1$, and $g(x) = h(x (1-p) + (1-x)p)$.
Observe that $f(x)$ and $g(x)$ are increasing differentiable functions in the region $[0,\frac 12]$. 
\begin{claim}
 \label{cl:decslope}
 For $\frac 16 \leq p \leq \frac{1}{2}$, the ratio of the derivatives $\frac{f^{(1)}(x)}{g^{(1)}(x)}$ is a decreasing function.
\end{claim}

The proof of this fact is found in the Appendix.

(Numerical simulations indicate that this is true for $p_{min} \leq p \leq \frac{1}{2}$ for $p_{min} \approx 0.05$, but for the purposes of establishing the inner bound clearly this region of $p$ suffices, as $\frac 16 \leq p_{max} \approx 0.184$). 

\begin{remark}
By combining Claim \ref{cl:decslope} and Corollary \ref{co:ineq} note that $h(p*f^{-1}(y))$ is convex in $y$, and this is very similar to Mrs. Gerber's Lemma \cite{wyz73}.
\end{remark}

Now let $s_{int}$ be defined according to

$$h(\frac{s_{int}}{2}) + h(\frac{1-s_{int}}{2}) = \sum_i u_i \left( h\big(\frac{s_i}{2}\big) + h\big(\frac{1-s_i}{2}\big)\right). $$

Then from Corollary \ref{co:ineq}, for $p_{min} \leq p \leq \frac{1}{2}$ we have
\begin{align*}
& h\big(\frac 12\big) - \sum_i u_i h(s_i (1-p) + (1-s_i)p) \\
& \quad \leq  h\big(\frac 12\big) - h(s_{int} (1-p) + (1-s_{int})p).
\end{align*}

This implies that the optimal auxiliary channel $U \to X$ is a BSC with a cross-over 
probability $s$ and $\p(U=0) = \frac{1}{2}$. Thus for $p_{max} \leq p \leq \frac{1}{2}$, 
 the boundary is characterized by the pair of points of the form,
\begin{align}
R_0 &= 1 - h(s (1-p) + (1-s)p),  \nonumber \\
R_1 &= \min\left\{\frac{1}{2} \left( h\big(\frac{s}{2}\big) + h\big(\frac{1-s}{2}\big) - 1 \right)\right.,\label{eq:ib}\\
&\quad \left. h\big(\frac{1}{4}\big) - \frac{3}{2} + h(s (1-p) + (1-s)p)\right\}, \nonumber
\end{align}
for $0 \leq s \leq \frac{1}{2}$. The second term in $R_1$ comes from taking into account the sum rate constraint,
$$ R_0 + R_1 \leq h\big(\frac{1}{4}\big) - \frac{1}{2}.$$

A simple calculation shows that for $p_o \leq p \leq \frac 12$ one can ignore the sum rate constraint, where $p_o = \frac{\sqrt{3}-1}{2\sqrt{3}} \approx 0.211$. This $p_o$ corresponds to the smallest value of $p$ where the convex region characterized by the pairs 
\begin{align*}
R_0 &= 1 - h(s (1-p) + (1-s)p),  \\
R_1 &= \frac{1}{2} \left( h\big(\frac{s}{2}\big) + h\big(\frac{1-s}{2}\big) - 1 \right).
\end{align*}
has a slope of $-1$ at the point $(R_0,R_1)=\left(0,h\big(\frac{1}{4}\big) - \frac{1}{2}\right)$.

Therefore the inner bound has three different expressions:
\begin{itemize}
\item $0 \leq p \leq p_{max}$: the inner bound reduces  to $R_0 + R_1 \leq h\big(\frac{1}{4}\big) - \frac{1}{2}$, 
\item $p_{max}\leq p \leq p_o$: the inner bound is given by equation \eqref{eq:ib} where all inequalities are necessary,
\item $p_o \leq p \leq \frac 12$: the inner bound is characterized by pair of points of the form
\begin{align*}
R_0 &= 1 - h(s (1-p) + (1-s)p),  \\
R_1 &= \frac{1}{2} \left( h\big(\frac{s}{2}\big) + h\big(\frac{1-s}{2}\big) - 1 \right).
\end{align*}
\end{itemize}

\subsection{Comparison with the outer bound}

To show that the outer bound gives a larger region, we produce a particular choice of the pair $(U_1,U_2,X)$.  Consider a $U_1, U_2$ defined as follows,
\begin{align*}
&\p(U_1=1) =\p(U_2=1) = u, \\
&\p(U_1=2)=\p(U_2=2) = 1-u, \\
&\p(X=0|U_1=1) = \p(X=1|U_2=1) = 1, \\
&\p(X=0|U_1=2) = \p(X=1|U_2=2) =  s,\\
\end{align*}
where $s= \frac{0.5 -u}{1-u}$ for $0\leq u \leq 0.5$. Existence of the triple $(U_1,U_2,X)$ is guaranteed by the consistent distribution on $X$. Substituting this choice into Bound \ref{bd:outer} we obtain {\em Region A} given by,
\begin{align*}
R_0 &\leq 1 - (1-u)h\big(s(1-p)+(1-s)p\big)-uh\big(p\big), \\
R_1 &\leq (1-u)h\big(\frac{s}{2}\big) - \frac 12 + u, \\
R_0 + R_1 & \leq h\big(\frac 14\big) - \frac  12.
\end{align*}

\begin{figure}[ht]
\begin{center}
\includegraphics[width=0.9\linewidth,angle=0]{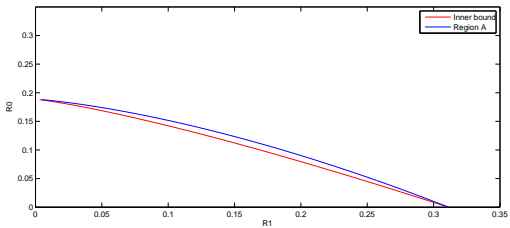}
\caption{Comparing Bound \ref{bd:inner} and Region A for $p = \frac 14$}
\label{fig:obib}
\end{center}
\end{figure}

Figure \ref{fig:obib} plots Region A and Bound \ref{bd:inner} for $p=\frac 14$. Observe that Region $A$ is larger than Bound \ref{bd:inner}, and hence the Bounds \ref{bd:inner} and \ref{bd:outer} do not match  for the 3-receiver channel shown in Figure \ref{fig:bsscbsc}. This implies the following corollary.

\begin{corollary}
 \label{co:differ}
There exists a class of channels, given in Figure \ref{fig:bsscbsc}, for which the inner and outer bounds (i.e. Bounds \ref{bd:inner} and \ref{bd:outer}) do not match.
\end{corollary}

\section{Revisiting outer bound}

We now show that the inner bound is tight for the channel shown in Figure \ref{fig:bsscbsc}.

Let $\pi: \{0,1\} \mapsto \{0,1\}; \pi(0)=1, \pi(1)=0$. 

Consider an $\epsilon$-codebook $\{x^n_{m_0,m_1}, 1 \leq m_0 \leq 2^{nR_0}, 1 \leq m_1 \leq 2^{nR_1}, \cA_{m_0,m_1} \subseteq \cY_1^n, \cB_{m_0,m_1} \subseteq \cY_2^n, \cC_{m_0} \subseteq \cY_3^n$\}, where the disjoint sets $\cA_{m_0,m_1}, \cB_{m_0,m_1}, \cC_{m_0}$ represent the decoding maps. From the skew symmetry of the channels $X \to (Y_1, Y_2)$ and the symmetry in channel $X \to Y_3$, it is clear that $\{\pi(x^n_{m_0,m_1}), 1 \leq m_0 \leq 2^{nR_0}, 1 \leq m_1 \leq 2^{nR_1}, \pi(\cB_{m_0,m_1}) \subseteq \cY_1^n, \pi(\cA_{m_0,m_1}) \subseteq \cY_2^n, \pi(\cC_{m_0}) \subseteq \cY_3^n$\} represents a valid $\epsilon$-codebook as well.

From these two codes, construct a new codebook (with error bounded by $\frac 12 + \epsilon$) and size $2^{nR_0} \times 2^{nR_1 + 1}$ as follows: The codewords are indexed by $x^n_{m_0,(m_1,b)}$ where $b=0,1$. When $b=0$ the codeword $x^n_{m_0,(m_1,b=0)}=x^n_{m_0,m_1}$ and when $b=1$, we have $x^n_{m_0,(m_1,b=1)}=\pi(x^n_{m_0,m_1})$. The decoding maps for this codebook are created as follows: If $y_1^n \in \cA_{m_0^1,m_1^1} \cap \pi(\cB_{m_0^2,m_1^2})$ then the receiver chooses one of the two message pairs $(m_0^1,m_1^1), (m_0^2,m_1^2)$ with equal probability. Otherwise it picks the message pair corresponding to the unique set $\cA_{m_0^1,m_1^1}$ or $\pi(\cB_{m_0^2,m_1^2})$ that it belongs to. A similar decoding strategy applies for receivers $Y_2$ and $Y_3$ as well.

The key feature is the symmetry of the codebook. If $x^n \in \mathbb{C}$ then $\pi(x^n) \in \mathbb{C}$ and correspond to the same message $M_0$.

Now observe that $H(M_0, M_1 | Y_1^n) \leq H(M_0, M_1,b | Y_1^n) \leq 1 +  H(M_0, M_1| Y_1^n,b) = 1 + n(R_0+R_1)\epsilon_n$. Therefore we obtain the same  outer bound (Bound \ref{bd:outer}) using
Fano's inequality and identification of the auxiliary random variables  as before.

In particular, the identifications of the auxiliary random variables remain the following:
$U_{1i} = (M_0, Y_{31}^{~i-1}, Y_{1i+1}^{~n})$ and $U_{2i} = (M_0, Y_{31}^{~i-1}, Y_{2i+1}^{~n})$.
Now for the skew-symmetric channels and a symmetric codebook observe that
{\footnotesize
\begin{align*}
&\p\Big(M_0=m_0,Y_{31}^{~i-1}=y_{31}^{i-1},Y_{1i+1}^{~n}=y_{1i+1}^{~n}, X_i=x_i\Big) \\
&  = \sum_{x_1^n\setminus x_i} \p\Big(M_0=m_0,X_1^{n}=x_1^n,   Y_{31}^{~i-1}=y_{31}^{~i-1},Y_{1i+1}^{~n}=y_{1i+1}^{~n} \Big) \\
&  \stackrel{(a)}{=} \sum_{x_1^n \setminus x_i} \p\Big(M_0=m_0,X_1^{n}=x_1^n \Big) \prod_{j=1}^{i-1} \p(Y_{3j}=y_{3j}|X_j=x_j)\\
& \qquad \times  \prod_{k=i+1}^{n} \p(Y_{1k}=y_{1k}|X_k=x_k) \\
&  \stackrel{(b)}{=} \sum_{x_1^{n}\setminus x_i} \p\Big(M_0=m_0,X_1^n = \pi(x_1^n) \Big) \prod_{j=1}^{i-1} \p(Y_{3j}=\pi(y_{3j})|X_j=\pi(x_j)) \\
& \qquad \times \prod_{k=i+1}^{n} \p(Y_{2k}=\pi(y_{1k})|X_k=\pi(x_k)) \\
&  =  \sum_{x_1^n\setminus x_i} \p\Big(M_0=m_0,X_1^{n}=\pi(x_1^n),   Y_{31}^{~i-1}=\pi(y_{31}^{~i-1}),Y_{2i+1}^{~n}=\pi(y_{1i+1}^{~n}) \Big) \\
&  \stackrel{(c)}{=} \p\Big(M_0=m_0,Y_{31}^{~i-1}=\pi(y_{31}^{i-1}),Y_{2i+1}^{~n}=\pi(y_{1i+1}^{~n}), X_i=\pi(x_i)\Big).
\end{align*}}

Here $(a)$ follows from the  discrete memoryless property of the channel; and $(b)$ follows from $(i)$ symmetry of the code, $(ii)$ symmetry of the channel $X \to Y_3$ with respect to $\pi(\cdot)$, and $(iii)$ the skew symmetry between receivers $Y_1, Y_2$ i.e. $$\p(Y_2 = \pi(y)|X=\pi(x)) = \p(Y_1=y|X=x);$$
and $(c)$ is a consequence of $\pi(\cdot)$ being a bijection.

Therefore the random variables $(U_1,X)$ and $(U_2,X)$ are identical up to re-labeling. Since the mutual information and entropy do not depend on the labeling, it follows that 
\begin{align*}
I(U_1;Y_3) &= I(U_2;Y_3) \\
I(X;Y_2|U_1) &= I(X;Y_2|U_2).
\end{align*}

\begin{remark}
This technique can be extended to other skew-symmetric channels as well, i.e one for which such a $\pi(\cdot)$ exists.
\end{remark}

Therefore we obtain the following revised outer bound.
\begin{bound}
 \label{bd:outer1}
The union over the set of rate pairs $(R_0,R_1)$ satisfying
\begin{align*}
 R_0 &\leq I(U_1;Y_3)\\
 R_0+R_1 &\leq \min\{I(U_1;Y_3)+I(X;Y_1|U_1), \\
 & \qquad I(U_1;Y_3)+I(X;Y_2|U_1)\}\\
 R_0 + R_1 &\leq \min\{I(X;Y_1),I(X;Y_2)\}
\end{align*}
over all possible choices of random variables $(U_1,X)$ such that $U_1 \to X \to (Y_1,Y_2,Y_3)$ forms a Markov chain
constitutes an outer bound for this channel.
\end{bound}

It is straightforward to see (using the boundary points) that Bound \ref{bd:outer1} matches the inner bound and forms the capacity region.

\section*{Acknowledgements}
The authors wish to thank Arvind Ramachandran for valuable suggestions and feedback on the contents of the paper.

\bibliographystyle{IEEEtran}
\bibliography{mybiblio}
%\bibliography{C:/Users/cnair/Desktop/documents/tex/mybiblio}

\appendix

\subsection{Proof of Claim \ref{cl:decslope}}
In this section we show that when $\frac{1}{6} \leq p \leq \frac{1}{2}$, the ratio
$\frac{f^{(1)}(x)}{g^{(1)}(x)}$ is a decreasing function of $x, x \in [0,\frac{1}{2}]$. Recalling the definitions, $f(x) = h(\frac{x}{2}) + h(\frac{1-x}{2}) - 1$, and $g(x) = h(x (1-p) + (1-x)p)$. As $f(x)$ and $g(x)$ are strictly increasing in  $x \in [0,\frac{1}{2}]$, it suffices to show that
\begin{equation}
 \frac{f^{(2)}(x)}{f^{(1)}(x)} \leq \frac{g^{(2)}(x)}{g^{(1)}(x)},
 \label{eq:apslope}
 \end{equation}
where $f^{(2)}(x), g^{(2)}(x)$ denote the second derivatives of the function.

Let $J(x) = \log \frac{1-x}{x}$, $U(x)=x(1-x)$ and $x*p = x(1-p) + p(1-x)$. Using this notation and substituting for the derivatives, \eqref{eq:apslope} reduces to showing
\begin{equation}
	\frac{J(x*p)U(x*p)}{1-2p} \geq \frac{2\Big( J\big(\frac x2\big) - J\big(\frac{1-x}{2}\big)\Big)}{\frac{1}{U(\frac{x}{2})} + \frac{1}{U(\frac{1-x}{2})}}.
	\label{eq:modapslope}
\end{equation}

Now observe that as $x \to \frac{1}{2}$ both $J(x*p)$ and $J\big(\frac x2\big) - J\big(\frac{1-x}{2}\big)$ tend to zero and all other terms remain positive. Thus we have an
equality at $x = \frac 12$. To show the inequality for $x \in [0,\frac 12]$ it  suffices to prove that the {\em derivative of the left hand side (L.H.S.)} of \eqref{eq:modapslope} is smaller than {\em derivative of the right hand side (R.H.S.)} of \eqref{eq:modapslope}.

The derivative of the L.H.S. is given by
$$ \frac{d}{dx}\frac{J(x*p)U(x*p)}{1-2p} = -1 + J(x*p)(1-2(x*p)). $$
Let us define $R(x)$ to be the derivative of the R.H.S., i.e.
$$ \frac{d}{dx} \frac{2\Big( J\big(\frac x2\big) - J\big(\frac{1-x}{2}\big)\Big)}{\frac{1}{U(\frac{x}{2})} + \frac{1}{U(\frac{1-x}{2})}} = R(x).$$

We wish to show that
\begin{equation}
	-1 + J(x*p)(1-2(x*p)) \leq R(x),
	\label{eq:condp}
\end{equation} 
for all $\frac 16 \leq p \leq \frac 12$ and $x \in [0,\frac{1}{2}]$. Given any $x \in [0,\frac 12]$, observe that $J(x*p)(1-2(x*p))$ is a decreasing function of $p$ for $0 \leq p \leq \frac 12$. Thus establishing \eqref{eq:condp} for $p = \frac{1}{6}$ suffices. 

\begin{figure}[ht]
\begin{center}
\includegraphics[width=\linewidth,angle=0]{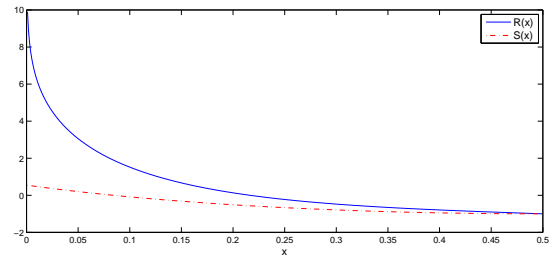}
\caption{Comparing $R(x)$ and $S(x)$}
\label{fig:comp}
\end{center}
\end{figure}

Let $S(x)=-1 + J(x*\frac{1}{6})(1-2(x*\frac{1}{6}))$. Figure \ref{fig:comp} plots 
$S(x)$ and  $R(x)$.

Thus we have $S(x) \leq R(x)$ for $0 \leq x \leq \frac{1}{2}$.  This completes the proof of Claim \ref{cl:decslope}.

\end{document}